\documentclass[12pt]{article}
\usepackage[dvips]{graphicx}
\begin{document}
\centerline{\large\bf Flavor changing neutrino interactions }
\vskip 0.4truecm
\centerline{\large\bf and CP violation in neutrino oscillations}
\baselineskip=8truemm
\vskip 0.8truecm
\centerline{Toshihiko Hattori$,^{a),}$\footnote{e-mail: 
hattori@ias.tokushima-u.ac.jp} \ Tsutom Hasuike$,^{b),}$\footnote{e-mail: 
hasuike@anan-nct.ac.jp} \ and \ Seiichi Wakaizumi$ ^{c),}$\footnote{e-mail: 
wakaizum@cue.tokushima-u.ac.jp}}
\vskip 0.5truecm
\centerline{\it $ ^{a)}$Institute of Theoretical Physics, University of Tokushima,
Tokushima 770-8502, Japan}
\centerline{\it $ ^{b)}$Department of Physics, Anan College of Technology,
Anan 774-0017, Japan}
\centerline{\it $ ^{c)}$Center for University Extension, University of 
Tokushima, Tokushima 770-8502, Japan}
\vskip 0.8truecm
\centerline{\bf Abstract}
\vskip 0.5truecm

We investigate the interference effects of non-standard neutrino-matter 
interactions (NSNI) with the mass-induced neutrino oscillations. 
The NSNI is composed of flavor-changing 
neutrino interactions (FCNI) and flavor-diagonal neutrino interactions (FDNI).
Both of the interactions are introduced in the $\nu_\mu -\nu_\tau$ sector 
and the $\nu_e -\nu_\mu$ sector in order to study their effects in $\nu_\mu
\to\nu_\tau$ and $\nu_\mu\to\nu_e$ oscillations, respectively. The FCNI effect 
proves to possibly dominate the CP violating effect and significantly survive 
as a fake CP violating effect in the 
neutrino energy region where the pure CP violating effect, ordinary matter 
effect and FDNI effect fall, for example, above 4 GeV at the baseline of 
$L=730$ km in the $\nu_\mu\to\nu_\tau$ oscillation for the maximum parameter 
values of FCNI and FDNI allowed by the atmospheric neutrino oscillation data. 
The FCNI and FDNI effects on CP violation in $\nu_\mu\to\nu_e$ 
oscillation are negligibly small due to the stringent constraints on FCNI 
from the bounds on lepton flavor violating processes and on FDNI from the 
limits on lepton universality violation. 
\newpage

\centerline{\large\bf I $\>$ Introduction}
\vskip 0.2truecm

In the framework of massive neutrinos and leptonic mixing, the atmospheric 
neutrino anomaly \cite{Hirata} is resolved by the $\nu_\mu\to\nu_\tau$ 
oscillation with nearly maximal mixing \cite{Fukuda} and the solar neutrino 
deficit \cite{Lande} is interpreted by the oscillation of $\nu_e$ into another 
neutrino state with large mixing angle solution \cite{Bahcall} 
in the Mikheyev-Smirnov-Wolfenstein (MSW) mechanism of neutrino interactions 
with matter \cite{Mikheyev} in the three-neutrino scheme, where the 
neutrino flavor eigenstates $\nu_\alpha\; (\alpha = e, \mu, \tau)$ are 
expressed by a superposition of their mass eigenstates $\nu_i\; (i = 1, 2, 3)$ 
with mass $m_i$  as follows: 
\begin{equation}
\nu_\alpha = \sum_{i=1}^3 U_{\alpha i}\, \nu_i ,    \label{shiki1}
\end{equation}
where $U$ is the $3\times 3$ unitary mixing matrix, which is called as 
Maki-Nakagawa-Sakata (MNS) matrix \cite{Maki}. The reactor experiment of 
search for $\bar{\nu_e}$ oscillations, CHOOZ, gives an upper limit on the 
element $U_{e3}$, which is very small as $|U_{e3}|<0.22$ \cite{CHOOZ}. 
The liquid scintillation experiment, LSND, claims a discovery of 
$\bar{\nu_\mu}\to\bar{\nu_e}$ oscillations \cite{LSND}, which requires 
a fourth sterile neutrino.

The above situation seems to convince us of a scheme of massive neutrinos 
and lepton mixing. In addition to this scheme, it is interesting to 
investigate non-standard neutrino-matter interactions in the 
neutrino oscillations in order to search for new physics beyond the standard 
model. The non-standard neutrino-matter interaction (NSNI) is 
originated from Wolfenstein's work \cite{Wolfenstein} and the flavor-changing 
neutrino-matter interaction (FCNI) and flavor-diagonal neutrino-matter 
interaction (FDNI) were used in order to study the solar neutrino problem, 
not by relying on the mass-induced neutrino oscillations 
\cite{Guzzo}\cite{Bergmann62}, and then by considering them as sub-leading 
effects to the standard mass-induced oscillations \cite{Bergmann}. For the 
atmospheric neutrino problem, the FDNI alone \cite{Ma} and both of 
the FCNI and FDNI were applied \cite{Garcia}. The mere NSNI, however, has 
proved not to be able to solve the atmospheric neutrino problem 
\cite{Bergmann00}\cite{Fornengo}. 

After that, the FCNI has been studied as sub-leading effects to the standard 
mass-induced neutrino oscillations by considering FCNI in $\nu_\mu -\nu_\tau$ 
sector and in $\nu_e -\nu_\tau$ sector \cite{Fornengo}\cite{Gago} and its 
detectability at a future neutrino factory is discussed in Refs.\cite{Gago}
\cite{Huber}\cite{Campanelli}. In Ref.19, it is shown that the FCNI dominates 
the $\nu_e$ oscillation probability at sufficiently high neutrino energies. 
Moreover, the NSNI with the netrino masses can induce some fake CP violating 
effect in matter in the long-baseline oscillation experiments \cite{Nunokawa}.
There are also combined analyses with new flavor-changing neutrino 
interactions occuring in the neutrino production and detection processes in 
addition to the above-mentioned non-standard interactions with matter during 
neutrino propagation, of which effects are enhanced by the interference with 
the ordinary weak interactions in the oscillation phenomena \cite{Gonzalez}.

In this paper, we analyze the non-standard neutrino-matter interaction 
effects (NSNI) to the CP violating effects in the neutrino oscillations by 
considering them in $\nu_\mu -\nu_\tau$ sector and $\nu_e -\nu_\mu$ sector 
in the three-neutrino scheme. The NSNI consists of the above-mentioned FCNI 
and flavor-diagonal neutrino-matter interactions (FDNI). The evolution 
equation of the neutrino flavor states is solved analytically by using 
Arafune-Koike-Sato's perturbative method \cite{Arafune} and we calculate the 
neutrino oscillation probability for the general $\nu_\alpha\to\nu_\beta$ 
oscillation. We apply this analytic expression of the probability to calculate 
the CP violating effects in $\nu_\mu\to\nu_\tau$ and $\nu_\mu\to\nu_e$ 
oscillations, {\it i.e.} the difference between the neutrino and the 
antineutrino oscillation probabilities. We find that the FCNI matter 
effect survives and dominates the CP violating effect in $\nu_\mu\to\nu_\tau$ 
oscillation even after both the 
pure CP violating effect due to the phase of $U$ and the fake CP violating 
ones due to the ordinary and FDNI matter effects fall. This shows that 
the non-standard flavor changing neutrino-matter interaction could be detected 
in the CP violating effect in $\nu_\mu\to\nu_\tau$ oscillation 
at the neutrino energies where both the pure CP violating effect and the 
ordinary matter effect become sufficiently small and undetectable. For 
$\nu_\mu\to\nu_e$ oscillation, the FCNI and FDNI effects on the CP violating 
effect are negligibly small due to the stringent constraints 
\cite{Bergmann62}\cite{Bergmann00}\cite{Bergmann59} on FCNI from the bounds 
on lepton flavor violating processes and on FDNI from the limits on lepton 
universality violation. 

The paper is organized as follows. In Sec. II the oscillation probability is 
derived by solving analytically the evolution equation of the neutrino flavor 
states with the non-standard neutrino-matter interactions in 
$\nu_\mu -\nu_\tau$ sector. The effect of the NSNI is studied in the CP 
violating effects in the $\nu_\mu\to\nu_\tau$ oscillation. In Sect. III 
the same will be done for the NSNI in $\nu_e -\nu_\mu$ sector and the effect 
of the NSNI is studied in the CP violating effects in the $\nu_\mu\to\nu_e$ 
oscillation. Section IV is devoted to the conclusions and discussions.
\vskip 0.7truecm

\centerline{\large\bf II $\>$ Oscillation probability and CP violating 
effect}
\centerline{\large\bf with NSNI in $\nu_\mu -\nu_\tau$ sector}
\vskip 0.2truecm

Here and in the next section we calculate the neutrino oscillation 
probabilities and CP violating effects with the inclusion of the NSNI in 
$\nu_\mu -\nu_\tau$ sector and in $\nu_e -\nu_\mu$ sector, respectively, in 
the three-neutrino scheme by solving analylitically the evolution equation for 
neutrino flavor states in the perturbative method.

If we consider the effect of NSNI in the $\nu_\mu -\nu_\tau$ sector, the 
evolution equation in matter is given as 
\begin{equation}
i\, \frac{d}{dx} \left (
\begin{array}{@{\,}c@{\,}}
\nu_e \\
\nu_\mu \\
\nu_\tau
\end{array}
\right ) = H \left (
\begin{array}{@{\,}c@{\,}}
\nu_e \\
\nu_\mu \\
\nu_\tau
\end{array}
\right ) ,     \label{shiki2}
\end{equation}
and \cite{Gago}
\begin{equation}
H = \frac{1}{2E} U \left(
\begin{array}{@{\,}ccc@{\,}}
0 & 0 & 0 \\
0 & \Delta m_{21}^2 & 0 \\
0 & 0 & \Delta m_{31}^2
\end{array}
\right) U^{\dagger} + \left(
\begin{array}{@{\,}ccc@{\,}}
V_e(x) & 0 & 0 \\
0 & 0 & \epsilon^f_{\mu\tau}V_f(x) \\
0 & \epsilon^{f*}_{\mu\tau}V_f(x) & \epsilon '^f_{\mu\tau}V_f(x)
\end{array}
\right)  ,     \label{shiki3}
\end{equation}
where $E$ is the neutrino energy, $\Delta m_{ij}^2 = m_i^2-m_j^2$, $m_i$ 
being the mass of $i$-th neutrino, $U$ is the MNS leptonic mixing matrix, 
$V_f(x)={\sqrt 2}G_Fn_f(x)$, $x$ being the position of the running neutrino, 
$\epsilon^f_{\mu\tau}V_f(x)$ is the flavor-changing $\nu_\mu +f\to\nu_\tau 
+f$ forward scattering amplitude due to the flavor-changing neutrino-matter 
interaction(FCNI), and $\epsilon '^f_{\mu\tau}V_f(x)$ is the difference 
between the flavor-diagonal $\nu_\mu -f$ and $\nu_\tau -f$ elastic forward 
scattering amplitudes due to the flavor-diagonal 
neutrino-matter interaction(FDNI), with $n_f(x)$ being the number density of 
the fermion $f (f=u, d, e)$ which induces such processes. In Eq.(3), 
$\epsilon^f_{\mu\tau}$ and $\epsilon '^f_{\mu\tau}$ are the phenomenological 
parameters which characterize the strength of FCNI and FDNI, respectively. 
The fermion number density $n_f(x)$ can be written in terms of the matter 
density $\rho$ as $n_f(x)=\rho (x)Y_f$, where $Y_f$ is the fraction of the 
fermion $f$ per nucleon, $\sim 1/2$ for electrons and $\sim 3/2$ for $u$ or 
$d$ quarks. In the following, we consider interactions only with either $u$ 
or $d$ quarks, since for the NSNI with electrons the same effects presented 
in this paper can be obtained simply by increasing the parameters 
$\epsilon^{u,d}_{\mu\tau}$ and $\epsilon '^{u,d}_{\mu\tau}$ by a factor 3.

For the evolution equation of the antineutrinos, the replacement of 
$U\to U^*, V_{e,f}(x)\to -V_{e,f}(x)$ and $\epsilon^f_{\mu\tau}\to 
\epsilon^{f*}_{\mu\tau}$ should be done in Eq.(3).

We use here Arafune-Koike-Sato's perturbative method to solve analytically 
the evolution equation \cite{Arafune}. The solution of Eq.(2) is given by
\begin{equation}
\nu (x) = S(x)\nu (0) ,    \label{shiki4}
\end{equation}
with 
\begin{equation}
S(x) = T{\rm exp}\left( -i\,\int_0^x ds H(s) \right)  ,   \label{shiki5}
\end{equation}
where 
\begin{equation}
\nu (x) = \left (
\begin{array}{@{\,}c@{\,}}
\nu_e(x) \\
\nu_\mu(x) \\
\nu_\tau(x)
\end{array}
\right )      \label{shiki6}
\end{equation}
and $T$ is the time ordering operator. In the following, the oscillation 
probability and the CP violating effect are calculated for the baseline of 
$L=300$ and 730 km so that we assume $n_f(x)$ and $\rho(x)$ to be independent 
of $x$. Then we have 
\begin{equation}
S(x) = e^{-i\,Hx}  .  \label{shiki7}
\end {equation}
The oscillation probability for $\nu_\alpha\to\nu_\beta$ at the distance $L$ 
from the neutrino production point is given in terms of $S$ in Eq.(5) as 
follows:
\begin{equation}
P(\nu_\alpha\to\nu_\beta ; L) = \left| S_{\beta\alpha}(L) \right|^2 .  
\label{shiki8}
\end{equation}

We express the Hamiltonian $H$ of Eq.(3) for simplicity as 
\begin{equation}
H = \frac{1}{2E} U \left(
\begin{array}{@{\,}ccc@{\,}}
0 & 0 & 0 \\
0 & \Delta m_{21}^2 & 0 \\
0 & 0 & \Delta m_{31}^2
\end{array}
\right) U^{\dagger} + \frac{1}{2E} \left(
\begin{array}{@{\,}ccc@{\,}}
a & 0 & 0 \\
0 & 0 & \epsilon b \\
0 & \epsilon^*b & \epsilon 'b
\end{array}
\right)  ,     \label{shiki9}
\end{equation}
where
\begin{eqnarray}
a &\equiv & 2EV_e = 2{\sqrt 2}G_Fn_eE = 7.60\times 10^{-5}\frac{\rho}
{[{\rm g}\,{\rm cm}^{-3}]}\frac{E}{[{\rm GeV}]}\;\; {\rm eV}^2 ,  
\label{shiki10}  \\
\epsilon b &\equiv & 2E\epsilon^f_{\mu\tau}V_f = 15.2\times 10^{-5}\epsilon 
Y_f\frac{\rho}{[{\rm g}\,{\rm cm}^{-3}]}\frac{E}{[{\rm GeV}]}\;\; {\rm eV}^2 ,
\label{shiki11}  \\
\epsilon 'b &\equiv & 2E\epsilon '^f_{\mu\tau}V_f = 15.2\times 10^{-5}
\epsilon 'Y_f\frac{\rho}{[{\rm g}\,{\rm cm}^{-3}]}\frac{E}{[{\rm GeV}]}
\;\; {\rm eV}^2  ,  \label{shiki12}
\end{eqnarray}
where $\epsilon^f_{\mu\tau}$ and $\epsilon '^f_{\mu\tau}$ are denoted as 
$\epsilon$ and $\epsilon '$, respectively, and $2EV_f$ is denoted 
as $b$, for brevity. 
In general, $\epsilon$ is complex and $\epsilon '$ is real from the 
hermeticity of the Hamiltonian. Since $\Delta m^2_{21} \ll \Delta m^2_{31}$ 
and $a, |\epsilon|b, |\epsilon '|b \ll \Delta m^2_{31}$ because of 
$|\epsilon|<0.02$ and $|\epsilon '|<0.05$ from the analysis of the 
atmospheric neutrino problem \cite{Fornengo}, we decompose $H$ of Eq.(9) as 
$H = H_0 + H_1$ with 
\begin{equation}
H_0 = \frac{1}{2E}U \left (
\begin{array}{@{\,}ccc@{\,}}
0 & 0 & 0 \\
0 & 0 & 0 \\
0 & 0 & \Delta m_{31}^2
\end{array}
\right ) U^{\dagger} ,        \label{shiki13}
\end{equation}
and 
\begin{equation}
H_1 = \frac{1}{2E} U \left(
\begin{array}{@{\,}ccc@{\,}}
0 & 0 & 0 \\
0 & \Delta m_{21}^2 & 0 \\
0 & 0 & 0
\end{array}
\right) U^{\dagger} + \frac{1}{2E} \left(
\begin{array}{@{\,}ccc@{\,}}
a & 0 & 0 \\
0 & 0 & \epsilon b \\
0 & \epsilon^* b & \epsilon 'b
\end{array}
\right)  ,     \label{shiki14}
\end{equation}
and treat $H_1$ as a perturbation and calculate Eq.(7) up to the first order 
in $\Delta m_{21}^2, a, \epsilon b$ and $\epsilon 'b$. Then, $S(x)$ is given 
by 
\begin{equation}
S(x) \simeq e^{-i\,H_0x} - i\, e^{-i\,H_0x}
\int_0^x ds H_1(s) ,   \label{shiki15}
\end{equation}
where $H_1(x) = e^{i\,H_0x}H_1\,e^{-i\,H_0x}$. 
The approximation in Eq.(15) requires 
\begin{equation}
\frac{\Delta m_{21}^2L}{2E} \ll 1 , \qquad 
\frac{|\epsilon|bL}{2E} \ll 1 , \qquad \frac{\epsilon 'bL}{2E} \ll 1 . 
\label{shiki16}
\end{equation}
The requirements of Eq.(16) are satisfied for 
$\Delta m_{21}^2=(10^{-5}-10^{-4})$ ${\rm eV}^2, E=1-20$ ${\rm GeV}, 
L = (300-730)$ km, $\rho = 3{\rm g/cm^3}, |\epsilon|\sim 0.02$ and 
$|\epsilon '|\sim 0.05$ as 
\begin{eqnarray}
\frac{\Delta m_{21}^2L}{2E} &\simeq & 4\times 10^{-4}-0.2 , \qquad  
\frac{|\epsilon|bL}{2E} \simeq (1-2.5)\times 10^{-2} ,  \nonumber  \\
\frac{|\epsilon '|bL}{2E} &\simeq & (2.5-6)\times 10^{-2} .   \label{shiki17}
\end{eqnarray}
Equation (16) also shows that the approximation becomes better as the energy 
$E$ increases. If we express $S_{\beta\alpha}(x)$ as 
\begin{equation}
S_{\beta\alpha}(x) = \delta_{\beta\alpha} + i\, T_{\beta\alpha}(x) , 
\label{shiki18}
\end{equation}
then $i\,T_{\beta\alpha}(x)$ is obtained as follows:
\begin{eqnarray}
i\,T_{\beta\alpha}(x) &=& -2\, i\, {\rm exp}\left( -i\,\frac{\Delta m^2_{31}x}
{4E} \right) \sin\left( \frac{\Delta m^2_{31}x}{4E} \right)
\Bigl[\> U_{\beta 3}U_{\alpha 3}^* \bigl\{ 1-\frac{a}{\Delta m^2_{31}}
(2|U_{e3}|^2    \nonumber  \\
& & {} -\delta_{\alpha e}-\delta_{\beta e})-i\,\frac{ax}{2E}|U_{e3}|^2
-\frac{\epsilon 'b}{\Delta m^2_{31}}(2|U_{\tau 3}|^2-\delta_{\alpha\tau}
-\delta_{\beta\tau})    \nonumber  \\
& & {} -i\,\frac{\epsilon 'bx}{2E}|U_{\tau 3}|^2 \bigr\}
-\frac{1}{\Delta m^2_{31}}\bigl\{ 4U^*_{\alpha 3}U_{\beta 3}{\rm Re}
(\epsilon bU^*_{\mu 3}U_{\tau 3})  \nonumber  \\
& & {} -\epsilon^*b(U^*_{\alpha 3}U_{\mu 3}\delta_{\beta\tau}
+U_{\beta 3}U^*_{\tau 3}\delta_{\alpha\mu})-\epsilon b(U^*_{\alpha 3}
U_{\tau 3}\delta_{\beta\mu}  \nonumber  \\
& & {} +U_{\beta 3}U^*_{\mu 3}\delta_{\alpha\tau}) \bigr\} -2\, i\, 
\frac{x}{2E}U^*_{\alpha 3}U_{\beta 3}{\rm Re}(\epsilon bU^*_{\mu 3}
U_{\tau 3})\> \Bigr]  \nonumber  \\
& & {} -i\,\frac{\Delta m^2_{31}x}{2E}\Bigl[\>\frac{\Delta m_{21}^2}
{\Delta m_{31}^2}U_{\beta 2}U_{\alpha 2}^*+\frac{a}{\Delta m^2_{31}}
\bigl\{ \delta_{\alpha e}\delta_{\beta e}+U_{\beta 3}U_{\alpha 3}^*
(2|U_{e3}|^2-\delta_{\alpha e}    \nonumber  \\
& & {} -\delta_{\beta e}) \bigr\}+\frac{\epsilon 'b}{\Delta m^2_{31}}\bigl\{ 
\delta_{\alpha\tau}\delta_{\beta\tau}+U_{\beta 3}U_{\alpha 3}^*
(2|U_{\tau 3}|^2-\delta_{\alpha\tau}-\delta_{\beta\tau}) \bigr\}  \nonumber \\
& & {} +\frac{1}{\Delta m^2_{31}}\bigl\{ \epsilon^*b\delta_{\beta\tau}
\delta_{\alpha\mu}+\epsilon b\delta_{\beta\mu}\delta_{\alpha\tau}
+4U^*_{\alpha 3}U_{\beta 3}{\rm Re}(\epsilon bU^*_{\mu 3}U_{\tau 3})  
\nonumber  \\
& & {} -\epsilon^*b(U^*_{\alpha 3}U_{\mu 3}\delta_{\beta\tau}
+U_{\beta 3}U^*_{\tau 3}\delta_{\alpha\mu})-\epsilon b(U^*_{\alpha 3}
U_{\tau 3}\delta_{\beta\mu}  \nonumber  \\
& & {} +U_{\beta 3}U^*_{\mu 3}\delta_{\alpha\tau}) \bigr\}\> \Bigr] .  
\label{shiki19}
\end{eqnarray}
We use Eq.(19) in Eq.(18) and calculate the oscillation probability for 
$\nu_\alpha\to\nu_\beta$ by Eq.(8). The complete expression of 
$P(\nu_\alpha\to\nu_\beta; L)$ with the NSNI in $\nu_\mu -\nu_\tau$ sector 
is given in the Appendix A. 

Now we will study the effects due to the NSNI in $\nu_\mu -\nu_\tau$ sector 
on the oscillation probability and the CP violating effect in 
$\nu_\mu\to\nu_\tau$ oscillation at the baseline of $L=730$ km. For the MNS 
mixing matrix $U$, we take the standard parametrization given by 
\begin{equation}
U = \left (
\begin{array}{@{\,}ccc@{\,}}
c_{12}c_{13} & s_{12}c_{13} & s_{13}e^{-i\,\delta} \\
-s_{12}c_{23}-c_{12}s_{23}s_{13}e^{i\,\delta} & c_{12}c_{23}-s_{12}s_{23}
s_{13}e^{i\,\delta} & s_{23}c_{13} \\
s_{12}s_{23}-c_{12}c_{23}s_{13}e^{i\,\delta} & -c_{12}s_{23}-s_{12}c_{23}
s_{13}e^{i\,\delta} & c_{23}c_{13}
\end{array}
\right )  ,        \label{shiki20}
\end{equation}
where $c_{ij}=\cos\theta_{ij}, s_{ij}=\sin\theta_{ij}$ and $\delta$ is the 
CP violating phase. From the Super-Kamiokande data for the atmospheric 
neutrino oscillation \cite{Fukuda}, $\sin^22\theta_{\rm atm}>0.82$ and 
$5\times 10^{-4}<\Delta m^2_{\rm atm}<6\times 10^{-3}$ ${\rm eV}^2$, 
we take in the following $\sin\theta_{23}=0.707$ for the mixing angle 
$s_{23}$ and $\Delta m^2_{31}=2.5\times 10^{-3}$ ${\rm eV}^2$ as a typical 
value. For $s_{12}$, we take $\sin\theta_{12}=0.54$ and $\Delta m^2_{21}
=7.3\times 10^{-5}$ ${\rm eV}^2$ from the presently most probable large-mixing 
angle solution (LMA) to the solar neutrino oscillation \cite{SKsol}. 
For $s_{13}$, we tentatively assume $\sin\theta_{13}=0.14$ from the CHOOZ 
data on $\bar{\nu_e}$ oscillation \cite{CHOOZ}, $\sin^2 2\theta_{\rm CHOOZ}
<0.18$ for $3\times 10^{-3}<\Delta m^2<1.0\times 10^{-2}$ ${\rm eV}^2$, 
which means $0<\sin\theta_{13}
<0.22$. For the phenomenological parameters of the NSNI, the constraints are 
derived to be $-0.03<\epsilon<0.02$ and $|\epsilon '|<0.05$ from the analyses 
of the atmospheric neutrino data by Fonengo et al. \cite{Fornengo}. They 
assumed $\epsilon$ to be real. Here we generally take $\epsilon$ to be 
complex. For the numerical calculations we take $|\epsilon|=0.01$ and 
$\epsilon '=\pm 0.02$, and the effects of the FCNI on CP violation have proved 
to be maximum at $\phi=0$ and $\pi$ for the phase of $\epsilon$, $\epsilon 
= |\epsilon|\, e^{i\phi}$.

In the following, we present the numerical results for the oscillation 
probabilities and CP violating effects in the $\nu_\mu\to\nu_\tau$ 
oscillation. The effect of NSNI in the $\nu_\mu -\nu_\tau$ sector is not so 
significant in the $\nu_\mu\to\nu_e$ and $\nu_e\to\nu_\tau$ oscillations as 
in the $\nu_\mu\to\nu_\tau$ oscillation. Fig.1 and Fig.2 show the $\nu_\mu
\to\nu_\tau$ oscillation probabilities for $|\epsilon|=0.01, 
\epsilon '=0.02, \delta=\pi/2, \Delta m^2_{31}>0$ for the neutrino energy 
range of $E=1-20$ GeV at the baseline of $L=730$ km for $\phi=0$ (Fig.1) and 
$\phi=\pi$ (Fig.2), respectively, where the solid line represents the 
oscillation probability including all the three matter effects, {\it i.e.} 
ordinary matter effect (denoted as $a$ in Eq.(14)), flavor-changing 
neutrino-matter interaction (FCNI, denoted as $\epsilon b$) and 
flavor-diagonal neutrino-matter interaction (FDNI, denoted as $\epsilon 'b$), 
and the dashed line represents the one without any matter effects. These two 
Figures show that the matter effects due to the NSNI are small for the 
oscillation probability.

Figs.3-6 show the CP violating effects in the $\nu_\mu\to\nu_\tau$ oscillation,
calculated as the difference between the neutrino and the antineutrino 
oscillation probabilities, for the energy range of $E=1-20$ GeV at the 
baseline of $L=730$ km. The phase of the MNS mixing matrix $U$ is taken as 
$\delta =\pi/2$. The dash-dotted line is the pure CP violating effect due 
to the phase of $U$. The long-dashed, short-dashed and dotted lines are the 
fake CP violating effects due to the ordinary, FCNI and FDNI matter effects, 
respectively. The solid line represents the total CP violating effect with 
the sum of the pure and fake ones. For Fig.3-6, we have taken 
$|\epsilon|=0.01,  |\epsilon '|=0.02$ 
and $|\Delta m^2_{31}|=2.5\times 10^{-3}$ ${\rm eV}^2$. In Fig.3, the phase 
of $\epsilon$ is taken as $\phi=0$ and $\epsilon '=+0.02$ $(>0)$ and 
$\Delta m^2_{31}>0$. In Fig.4, $\phi=\pi, \epsilon '=+0.02$ and 
$\Delta m^2_{31}>0$. In Fig.5, $\phi=0, \epsilon '>0$ and $\Delta m^2_{31}<0$, 
and in Fig.6, $\phi=\pi, \epsilon '>0$ and 
$\Delta m^2_{31}<0$. As can be seen from these Figures, the FCNI matter effect 
dominates the CP violating effect in the whole range of $E=1-20$ GeV, and all 
the other contributions of pure CP violation, ordinary matter effect and FDNI 
effect rapidly fall around 4 GeV and the FCNI effect survives significantly 
above this energy. When the sign of $\Delta m^2_{31}$ is changed, the 
contributions of all the matter effects including FCNI and FDNI change the 
sign, as can be seen from the comparison of Fig.3 and Fig.5 and from Fig.4 
and Fig.6. 

Next we will show how much the predicted CP violation depends on the 
oscillation parameters in their allowed ranges. Fig.7 and Fig.8 give the 
dependence of the total CP violating effect in $\nu_\mu\to\nu_\tau$ 
oscillation at $L=730$ km on the mass-squared differences $\Delta m^2_{31}$ 
and $\Delta m^2_{21}$ for $\Delta m^2_{31}=1.5\times 10^{-3}, 2.5\times 
10^{-3}, 3.5\times 10^{-3}$ ${\rm eV}^2$ and $\Delta m^2_{21}=6\times 10^{-5}, 
7\times 10^{-5}, 8\times 10^{-5}$ ${\rm eV}^2$, respectively. The dependence 
on $\Delta m^2_{31}$ is large and that on $\Delta m^2_{21}$ is very small, 
because the baseline of $L=730$ km corresponds to the atmospheric neutrino 
mass scale. The dash-dotted curve in both Figures represents the pure CP 
violating effect without the FCNI and FDNI. Figs. 9, 10 and 11 give the 
dependence of the total CP violating effect on the mixing angles $s_{23}, 
s_{12}$, and $s_{13}$ for $s_{23}=0.55, 0.60, 0.65, s_{12}=0.50, 0.55, 0.60$ 
and $s_{13}=0.05, 0.10, 0.15$, respectively. The dependence on the mixing 
angles is evidently small. Again, the dash-dotted curve represents the pure 
CP violating effect without the FCNI and FDNI. These Figures 7-11 show that 
the new CP violating effects are significantly larger than the pure CP 
violating one coming from the standard model. 
\vskip 0.7truecm

%\newpage
\centerline{\large\bf III $\>$ NSNI in $\nu_e -\nu_\mu$ sector}
\vskip 0.2truecm

In this section, we consider the NSNI in $\nu_e -\nu_\mu$ sector and study 
its effect on the CP violating effect in 
$\nu_\mu\to\nu_e$ oscillation at the baselines of $L=300$ km.

The Hamiltonian of the evolution equation of the flavor neutrino states is 
given as
\begin{equation}
H = \frac{1}{2E} U \left(
\begin{array}{@{\,}ccc@{\,}}
0 & 0 & 0 \\
0 & \Delta m_{21}^2 & 0 \\
0 & 0 & \Delta m_{31}^2
\end{array}
\right) U^{\dagger} + \left(
\begin{array}{@{\,}ccc@{\,}}
V_e(x) & \epsilon^f_{e\mu}V_f(x) & 0 \\
\epsilon^{f*}_{e\mu}V_f(x) & \epsilon '^f_{e\mu}V_f(x) & 0 \\
0 & 0 & 0
\end{array}
\right)  ,     \label{shiki21}
\end{equation}
where $\epsilon^f_{e\mu}V_f(x)$ is the flavor-changing $\nu_e +f\to\nu_\mu 
+f$ forward scattering amplitude due to the FCNI and 
$\epsilon '^f_{e\mu}V_f(x)$ is the difference between the flavor-diagonal 
$\nu_e -f$ and $\nu_\mu -f$ elastic 
forward scattering amplitudes due to the FDNI. As in the previous section, 
we assume the matter density to be constant for the baseline of $L=300$ km 
and reexpress Eq.(21) as 
\begin{equation}
H = \frac{1}{2E} U \left(
\begin{array}{@{\,}ccc@{\,}}
0 & 0 & 0 \\
0 & \Delta m_{21}^2 & 0 \\
0 & 0 & \Delta m_{31}^2
\end{array}
\right) U^{\dagger} + \frac{1}{2E} \left(
\begin{array}{@{\,}ccc@{\,}}
a & \eta b & 0 \\
\eta^*b & \eta 'b & 0 \\
0 & 0 & 0
\end{array}
\right)  ,     \label{shiki22}
\end{equation}
where $a$ is the same as in Eq.(10) and 
\begin{eqnarray}
\eta b &\equiv & 2E\epsilon^f_{e\mu}V_f = 15.2\times 10^{-5}\eta 
Y_f\frac{\rho}{[{\rm g}\,{\rm cm}^{-3}]}\frac{E}{[{\rm GeV}]}\;\; {\rm eV}^2 ,
\label{shiki23}  \\
\eta 'b &\equiv & 2E\epsilon '^f_{e\mu}V_f = 15.2\times 10^{-5}\eta '
Y_f\frac{\rho}{[{\rm g}\,{\rm cm}^{-3}]}\frac{E}{[{\rm GeV}]}\;\; {\rm eV}^2 ,
\label{shiki24}
\end{eqnarray}
where $\epsilon^f_{e\mu}$ and $\epsilon '^f_{e\mu}$ are denoted as $\eta$ and 
$\eta '$, respectively, and $2EV_f$ is denoted as $b$, for brevity. 
In general, $\eta$ is complex and $\eta '$ is real. The experimental limits on 
various lepton flavor violating processes and $SU(2)_L$ breaking effects give 
a stringent constraint on the FCNI parameter as $|\eta| < 7\times 10^{-5}$ 
\cite{Bergmann62}\cite{Bergmann00}\cite{Bergmann59} and the upper bounds on 
lepton universality violation give a constraint on the FDNI parameter as 
$|\eta '| < 0.1$ \cite{Bergmann62}\cite{Bergmann00}. So, we can decompose $H$ 
of Eq.(22) as $H=H_0+H_1$ with
\begin{equation}
H_1 = \frac{1}{2E} U \left(
\begin{array}{@{\,}ccc@{\,}}
0 & 0 & 0 \\
0 & \Delta m_{21}^2 & 0 \\
0 & 0 & 0
\end{array}
\right) U^{\dagger} + \frac{1}{2E} \left(
\begin{array}{@{\,}ccc@{\,}}
a & \eta b & 0 \\
\eta^*b & \eta 'b & 0 \\
0 & 0 & 0
\end{array}
\right)  ,     \label{shiki25}
\end{equation}
and treat $H_1$ as a perturbation. $H_0$ is the same as in Eq.(13). 
In the same way as for the NSNI in $\nu_\mu -\nu_\tau$ sector, we calculate 
$S(x)$ of Eq.(7) up to the first order in $\Delta m^2_{21}, a, \eta b$ and 
$\eta 'b$ to obtain the expression of $i\,T_{\beta\alpha}(x)$ in Eq.(18):
\begin{eqnarray}
i\,T_{\beta\alpha}(x) &=& -2\, i\, {\rm exp}\left( -i\,\frac{\Delta m^2_{31}x}
{4E} \right) \sin\left( \frac{\Delta m^2_{31}x}{4E} \right)
\Bigl[\> U_{\beta 3}U_{\alpha 3}^* \bigl\{ 1-\frac{a}{\Delta m^2_{31}}
(2|U_{e3}|^2    \nonumber  \\
& & {} -\delta_{\alpha e}-\delta_{\beta e})-i\,\frac{ax}{2E}|U_{e3}|^2
-\frac{\eta 'b}{\Delta m^2_{31}}(2|U_{\mu 3}|^2-\delta_{\alpha\mu}
-\delta_{\beta\mu})    \nonumber  \\
& & {} -i\,\frac{\eta 'bx}{2E}|U_{\mu 3}|^2 \bigr\}
-\frac{1}{\Delta m^2_{31}}\bigl\{ 4U^*_{\alpha 3}U_{\beta 3}{\rm Re}
(\eta bU^*_{e3}U_{\mu 3})  \nonumber  \\
& & {} -\eta^*b(U^*_{\alpha 3}U_{e3}\delta_{\beta\mu}
+U_{\beta 3}U^*_{\mu 3}\delta_{\alpha e})-\eta b(U^*_{\alpha 3}
U_{\mu 3}\delta_{\beta e}  \nonumber  \\
& & {} +U_{\beta 3}U^*_{e3}\delta_{\alpha\mu}) \bigr\} -2\, i\, 
\frac{x}{2E}U^*_{\alpha 3}U_{\beta 3}{\rm Re}(\eta bU^*_{e3}
U_{\mu 3})\> \Bigr]  \nonumber  \\
& & {} -i\,\frac{\Delta m^2_{31}x}{2E}\Bigl[\> \frac{\Delta m_{21}^2}
{\Delta m_{31}^2}U_{\beta 2}U_{\alpha 2}^*+\frac{a}{\Delta m^2_{31}}
\bigl\{ \delta_{\alpha e}\delta_{\beta e}+U_{\beta 3}U_{\alpha 3}^*
(2|U_{e3}|^2-\delta_{\alpha e}    \nonumber  \\
& & {} -\delta_{\beta e}) \bigr\} +\frac{\eta 'b}{\Delta m~2_{31}}\bigl\{ 
\delta_{\alpha\mu}\delta_{\beta\mu}+U_{\beta 3}U_{\alpha 3}^*
(2|U_{\mu 3}|^2-\delta_{\alpha\mu}-\delta_{\beta\mu}) \bigr\}  \nonumber \\
& & {} +\frac{1}{\Delta m^2_{31}}\bigl\{ \eta^*b\delta_{\beta\mu}
\delta_{\alpha e}+\eta b\delta_{\beta e}\delta_{\alpha\mu}
+4U^*_{\alpha 3}U_{\beta 3}{\rm Re}(\eta bU^*_{e3}U_{\mu 3})  
\nonumber  \\
& & {} -\eta^*b(U^*_{\alpha 3}U_{e3}\delta_{\beta\mu}
+U_{\beta 3}U^*_{\mu 3}\delta_{\alpha e})-\eta b(U^*_{\alpha 3}
U_{\mu 3}\delta_{\beta e}  \nonumber  \\
& & {} +U_{\beta 3}U^*_{e3}\delta_{\alpha\mu}) \bigr\}\> \Bigr] .  
\label{shiki26}
\end{eqnarray}
Using Eq.(26) in Eq.(18), we calculate the oscillation probability for 
$\nu_\alpha\to\nu_\beta$ by Eq.(8). The complete expression of 
$P(\nu_\alpha\to\nu_\beta; L)$ with the NSNI in $\nu_e -\nu_\mu$ sector 
is given in the Appendix B. 

Now we will study the effects due to the NSNI in $\nu_e -\nu_\mu$ sector 
on the CP violating effect in $\nu_\mu\to\nu_e$ oscillation at the baseline 
of $L=300$ km. The values of the parameters $\Delta m^2_{21}, \Delta m^2_{31}, 
s_{23}, s_{12}$ and $s_{13}$ are taken to be the same as in the previous 
section. The effect of the FCNI on CP violation has proved to be maximum 
at $\gamma=\pi/2$ and $3\pi/2$ for the phase of $\eta, 
\eta=|\eta|{\rm e}^{i\,\gamma}$. We present the numerical results for the 
CP violating effects in the $\nu_{\mu}\to\nu_e$ oscillation for the 
parameters satisfying the constraints. In Fig.12 
we show the CP violating effect in $\nu_{\mu}\to\nu_e$ oscillation 
for $|\eta|=7\times 10^{-5}, \gamma=\pi/2, \eta '=0.01$ 
in the neutrino energy range of $E=0.3-2$ GeV at the baseline of $L=300$ km 
for $\Delta m^2_{31}>0$ and $\delta=\pi/2$ for the phase of $U$. 
The dash-dotted line represents the pure CP violating effect due to the phase 
of $U$. The long-dashed and short-dashed lines are the fake 
CP violating effects due to the ordinary and FCNI matter effects, 
respectively. The solid line represents the total CP violating effect, 
which is the sum of the pure and fake ones. The FCNI effect 
turns out to be negligibly small, so that the total CP violating effect is 
just given by the pure CP violating effect and the ordinary matter effect. 
The FDNI effect is negligibly small just as the FCNI effect and is not shown 
in Fig.12.
\vskip 0.7truecm

\centerline{\large\bf IV $\>$ Conclusions and discussions}
\vskip 0.2truecm

In this paper we have studied the effects of non-standard neutrino-matter 
interactions on the standard mass-induced oscillation probabilities and 
the CP violating effects. The non-standard interactions are introduced 
in the flavor-changing neutrino interaction (FCNI) and the flavor-diagonal 
one (FDNI) as sub-leading effects in the $\nu_\mu -\nu_\tau$ sector and 
$\nu_e -\nu_\mu$ sector in order to investigate their effects in $\nu_\mu\to 
\nu_\tau$ and $\nu_\mu\to\nu_e$ oscillations, respectively, at the baselines 
of $L=730$ and 300 km. By using the values of FCNI and FDNI parameters 
allowed by the atmospheric neutrino oscillation analyses, in 
$\nu_\mu\to\nu_\tau$ oscillation at $L=730$ km the FCNI contribution 
has proved to dominate the CP violating effect and survive significantly in 
the energy region where all the others of pure CP violating effect, ordinary 
matter effect and FDNI contribution fall. In $\nu_\mu\to\nu_e$ oscillation 
at $L=300$ km, the FCNI and FDNI contributions are negligibly small due to 
the stringent constraints on FCNI from the bounds on various lepton flavor 
violating processes and on FDNI from the limits on lepton universality 
violation.

These results show that the non-standard neutrino-matter interations, 
especially the flavor-changing neutrino interaction, could be detected 
in the CP violating effect in $\nu_\mu\to\nu_\tau$ oscillation at the 
neutrino energies where both the pure CP violating effect and the ordinary 
matter effect fall, for example, above 4 GeV at $L=730$ km.\footnote[1]
{After the completion of 
this work, Prof. Branco informed us that they studied the effect of the 
addition of a new isosinglet charged lepton inspired by extra dimensions to 
the standard spectrum on the CP asymmetries in neutrino oscillations in 
$\nu_e-\nu_\mu$ and $\nu_\mu-\nu_\tau$ channels and its detectability 
at future neutrino factories. \cite{Branco}}
%\vskip 0.7truecm

\newpage
\centerline{\large\bf Appendix A:  Oscillation probability}
\centerline{\large\bf with NSNI in $\nu_\mu -\nu_\tau$ sector}
\vskip 0.2truecm

Here we present the oscillation probability of Eq.(8) with Eq.(19) taken in 
Eq.(18). 
\begin{eqnarray}
&P&(\nu_\alpha\to\nu_\beta; L)    \nonumber  \\
&=& \delta_{\beta\alpha}\Bigl[\> 1-4\sin^2\left( \frac{\Delta m^2_{31}L}{4E} 
\right) \Bigl\{ |U_{\alpha 3}|^2\bigl[\> 1-2\frac{a}{\Delta m^2_{31}}
(|U_{e3}|^2-\delta_{\alpha e})    \nonumber  \\
& & {} -2\frac{\epsilon 'b}{\Delta m^2_{31}}(|U_{\tau 3}|^2
-\delta_{\alpha\tau})\> \bigr]-\frac{2}{\Delta m^2_{31}}{\rm Re}(\epsilon b
U^*_{\mu 3}U_{\tau 3})(2|U_{\alpha 3}|^2-\delta_{\alpha\mu}
-\delta_{\alpha\tau}) \Bigr\}   \nonumber  \\
& & {} -2\sin\left( \frac{\Delta m^2_{31}L}{2E} \right)|U_{\alpha 3}|^2
\Bigl\{ \frac{aL}{2E}|U_{e3}|^2+2\frac{L}{2E}{\rm Re}(\epsilon bU^*_{\mu 3}
U_{\tau 3})+\frac{\epsilon 'bL}{2E}|U_{\tau 3}|^2 \Bigr\} \>\Bigr]    
\nonumber  \\
& & {} +4\sin^2\left( \frac{\Delta m^2_{31}L}{4E}\right) \Bigl[\> 
|U_{\alpha 3}|^2|U_{\beta 3}|^2 \bigl\{1-2\frac{a}{\Delta m^2_{31}}
(2|U_{e3}|^2-\delta_{\alpha e}-\delta_{\beta e})    \nonumber  \\
& & {} -2\frac{\epsilon 'b}{\Delta m^2_{31}}(2|U_{\tau 3}|^2
-\delta_{\alpha\tau}-\delta_{\beta\tau}) \bigr\}-\frac{2}{\Delta m^2_{31}}
{\rm Re}(\epsilon bU^*_{\mu 3}U_{\tau 3})\bigl\{ 4|U_{\alpha 3}|^2
|U_{\beta 3}|^2   \nonumber  \\
& & {} -|U_{\alpha 3}|^2(\delta_{\beta\mu}+\delta_{\beta\tau})-|U_{\beta 3}|^2
(\delta_{\alpha\mu}+\delta_{\alpha\tau})\bigr\} \>\Bigr]
+2\frac{\Delta m^2_{31}L}{2E}\sin\left( \frac{\Delta m^2_{31}L}{2E} \right)
    \nonumber  \\
& & {} \times \Bigl[\> \frac{\Delta m^2_{21}}{\Delta m^2_{31}}
{\rm Re}(U_{\beta 3}U_{\beta 2}^*U_{\alpha 3}^*U_{\alpha 2})
+\frac{a}{\Delta m^2_{31}}\bigl\{ \delta_{\alpha e}\delta_{\beta e}|U_{e3}|^2
+|U_{\alpha 3}|^2|U_{\beta 3}|^2(2|U_{e3}|^2   \nonumber  \\
& & {} -\delta_{\alpha e}-\delta_{\beta e})\bigr\}
+\frac{\epsilon 'b}{\Delta m^2_{31}}\bigl\{ \delta_{\alpha\tau}
\delta_{\beta\tau}|U_{\tau 3}|^2 +|U_{\alpha 3}|^2|U_{\beta 3}|^2
(2|U_{\tau 3}|^2-\delta_{\alpha\tau}-\delta_{\beta\tau}) \bigr\} \nonumber  \\
& & {} +\frac{1}{\Delta m^2_{31}}{\rm Re}(\epsilon bU^*_{\mu 3}U_{\tau 3})
\bigl\{ \delta_{\alpha\mu}\delta_{\beta\tau}+\delta_{\alpha\tau}
\delta_{\beta\mu}+4|U_{\alpha 3}|^2|U_{\beta 3}|^2   \nonumber  \\
& & {} -|U_{\alpha 3}|^2(\delta_{\beta\mu}+\delta_{\beta\tau})
-|U_{\beta 3}|^2(\delta_{\alpha\mu}+\delta_{\alpha\tau}) \bigr\} 
\>\Bigr]+4\frac{\Delta m^2_{31}L}{2E}\sin^2\left( \frac{\Delta m^2_{31}L}
{4E} \right)   \nonumber  \\
& & {} \times \Bigl[\> \frac{\Delta m^2_{21}}{\Delta m^2_{31}}
{\rm Im}(U_{\beta 3}U_{\beta 2}^*U_{\alpha 3}^*U_{\alpha 2})
+\frac{1}{\Delta m^2_{31}}{\rm Im}(\epsilon bU^*_{\mu 3}U_{\tau 3})
\bigl\{ \delta_{\alpha\mu}\delta_{\beta\tau}
-\delta_{\alpha\tau}\delta_{\beta\mu}   \nonumber  \\
& & {} +|U_{\alpha 3}|^2(\delta_{\beta\mu}-\delta_{\beta\tau})
-|U_{\beta 3}|^2(\delta_{\alpha\mu}-\delta_{\alpha\tau}) \bigr\} 
\>\Bigr]. \qquad  \qquad  \qquad  \qquad  \qquad ({\rm A1}) \nonumber
\end{eqnarray}
\vskip 0.7truecm

\newpage
\centerline{\large\bf Appendix B:  Oscillation probability}
\centerline{\large\bf with NSNI in $\nu_e -\nu_\mu$ sector}
\vskip 0.2truecm

Here we present the oscillation probability of Eq.(8) with Eq.(26) taken in 
Eq.(18). 
\begin{eqnarray}
&P&(\nu_\alpha\to\nu_\beta; L)    \nonumber  \\
&=& \delta_{\beta\alpha}\Bigl[\> 1-4\sin^2\left( \frac{\Delta m^2_{31}L}{4E} 
\right) \Bigl\{ |U_{\alpha 3}|^2\bigl[\> 1-2\frac{a}{\Delta m^2_{31}}
(|U_{e3}|^2-\delta_{\alpha e})    \nonumber  \\
& & {} -2\frac{\eta 'b}{\Delta m^2_{31}}(|U_{\mu 3}|^2
-\delta_{\alpha\mu})\> \bigr]-\frac{2}{\Delta m^2_{31}}{\rm Re}(\eta b
U^*_{e3}U_{\mu 3})(2|U_{\alpha 3}|^2-\delta_{\alpha\mu}
-\delta_{\alpha e}) \Bigr\}   \nonumber  \\
& & {} -2\sin\left( \frac{\Delta m^2_{31}L}{2E} \right)|U_{\alpha 3}|^2
\Bigl\{ \frac{aL}{2E}|U_{e3}|^2+2\frac{L}{2E}{\rm Re}(\eta bU^*_{e3}
U_{\mu 3})+\frac{\eta 'bL}{2E}|U_{\mu 3}|^2 \Bigr\} \>\Bigr]    \nonumber  \\
& & {} +4\sin^2\left( \frac{\Delta m^2_{31}L}{4E }\right) \Bigl[\> 
|U_{\alpha 3}|^2|U_{\beta 3}|^2 \bigl\{ 1-2\frac{a}{\Delta m^2_{31}}
(2|U_{e3}|^2-\delta_{\alpha e}-\delta_{\beta e})    \nonumber  \\
& & {} -2\frac{\eta 'b}{\Delta m^2_{31}}(2|U_{\mu 3}|^2
-\delta_{\alpha\mu}-\delta_{\beta\mu}) \bigr\}-\frac{2}{\Delta m^2_{31}}
{\rm Re}(\eta bU^*_{e3}U_{\mu 3})\bigl\{ 4|U_{\alpha 3}|^2|U_{\beta 3}|^2   
\nonumber  \\
& & {} -|U_{\alpha 3}|^2(\delta_{\beta e}+\delta_{\beta\mu})-|U_{\beta 3}|^2
(\delta_{\alpha e}+\delta_{\alpha\mu})\bigr\} \>\Bigr]
+2\frac{\Delta m^2_{31}L}{2E}\sin\left( \frac{\Delta m^2_{31}L}{2E} 
\right)    \nonumber  \\
& & {} \times \Bigl[\> \frac{\Delta m^2_{21}}{\Delta m^2_{31}}
{\rm Re}(U_{\beta 3}U_{\beta 2}^*U_{\alpha 3}^*U_{\alpha 2})
+\frac{a}{\Delta m^2_{31}}\bigl\{\delta_{\alpha e}\delta_{\beta e}|U_{e3}|^2
+|U_{\alpha 3}|^2|U_{\beta 3}|^2(2|U_{e3}|^2  \nonumber  \\
& & {} -\delta_{\alpha e}-\delta_{\beta e})\bigr\}
+\frac{\eta 'b}{\Delta m^2_{31}}\bigl\{ \delta_{\alpha\mu}
\delta_{\beta\mu}|U_{\mu 3}|^2 +|U_{\alpha 3}|^2|U_{\beta 3}|^2
(2|U_{\mu 3}|^2-\delta_{\alpha\mu}-\delta_{\beta\mu}) \bigr\}  \nonumber  \\
& & {} +\frac{1}{\Delta m^2_{31}}{\rm Re}(\eta bU^*_{e3}U_{\mu 3})
\bigl\{ \delta_{\alpha e}\delta_{\beta\mu}+\delta_{\alpha\mu}\delta_{\beta e}
+4|U_{\alpha 3}|^2|U_{\beta 3}|^2   \nonumber  \\
& & {} -|U_{\alpha 3}|^2(\delta_{\beta e}+\delta_{\beta\mu})
-|U_{\beta 3}|^2(\delta_{\alpha e}+\delta_{\alpha\mu}) \bigr\} \>\Bigr]
+4\frac{\Delta m^2_{31}L}{2E}\sin^2\left( \frac{\Delta m^2_{31}L}
{4E} \right)   \nonumber  \\
& & {} \times \Bigl[\> \frac{\Delta m^2_{21}}{\Delta m^2_{31}}
{\rm Im}(U_{\beta 3}U_{\beta 2}^*U_{\alpha 3}^*U_{\alpha 2})
+\frac{1}{\Delta m^2_{31}}{\rm Im}(\eta bU^*_{e3}U_{\mu 3})
\bigl\{ \delta_{\alpha e}\delta_{\beta\mu}
-\delta_{\alpha\mu}\delta_{\beta e}     \nonumber  \\
& & {} +|U_{\alpha 3}|^2(\delta_{\beta e}-\delta_{\beta\mu})
-|U_{\beta 3}|^2(\delta_{\alpha e}-\delta_{\alpha\mu}) \bigr\} 
\>\Bigr]. \qquad  \qquad  \qquad  \qquad  \qquad ({\rm B1}) \nonumber
\end{eqnarray}

\newpage
\begin{figure}[tbp]
\begin{center}
\includegraphics[width=13cm]{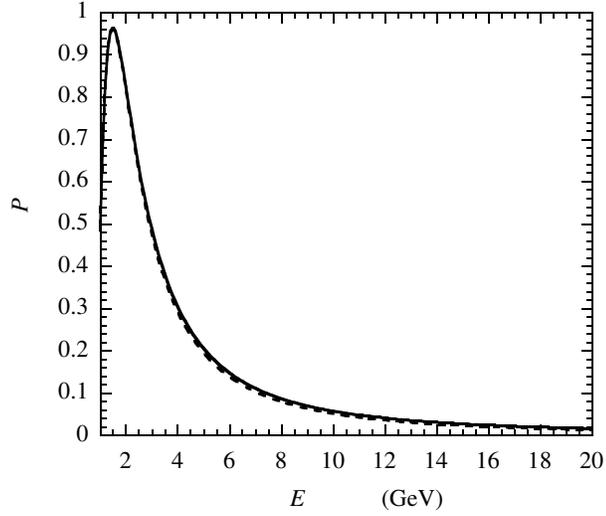}
%\vspace{10mm}
\caption{$\nu_\mu\to\nu_\tau$ oscillation probability for the neutrino 
energy range $E=1-20$ GeV at the baseline of $L=730$ km, with the NSNI in 
$\nu_\mu -\nu_\tau$ sector. The solid line is the one with all of the 
ordinary, FCNI and FDNI matter effects included, for $|\epsilon|=0.01, 
\epsilon '=0.02, \phi=0$ for the phase of $\epsilon$, $\delta=\pi/2$ for 
the phase of $U, \Delta m^2_{21}=7.3\times 10^{-5}$ ${\rm eV}^2$, and 
$\Delta m^2_{31}=2.5\times 10^{-3}{\rm eV}^2$ $(>0)$. The dashed line is 
the one without any matter effects.}
\end{center}
\end{figure}

\newpage
\begin{figure}[tbp]
\begin{center}
\includegraphics[width=13cm]{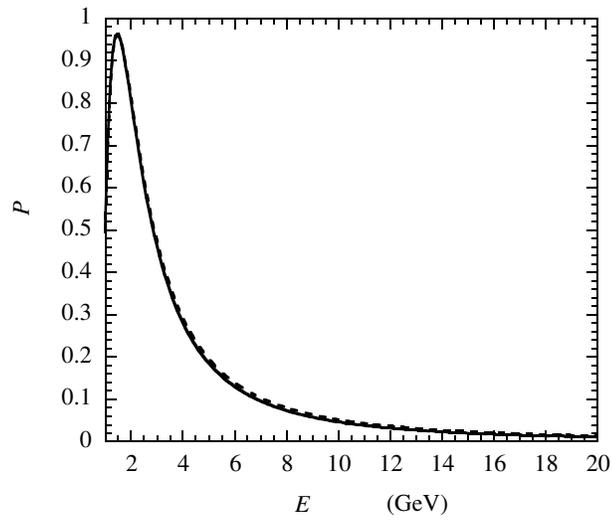}
%\vspace{10mm}
\caption{$\nu_\mu\to\nu_\tau$ oscillation probability for the neutrino 
energy range $E=1-20$ GeV at $L=730$ km with all of the ordinary, FCNI and 
FDNI matter effects included (solid line) and without any matter effects 
(dashed line). The parameter values are the same as in Fig.1 except for 
$\phi=\pi$.}
\end{center}
\end{figure}

\newpage
\begin{figure}[tbp]
\begin{center}
\includegraphics[width=13cm]{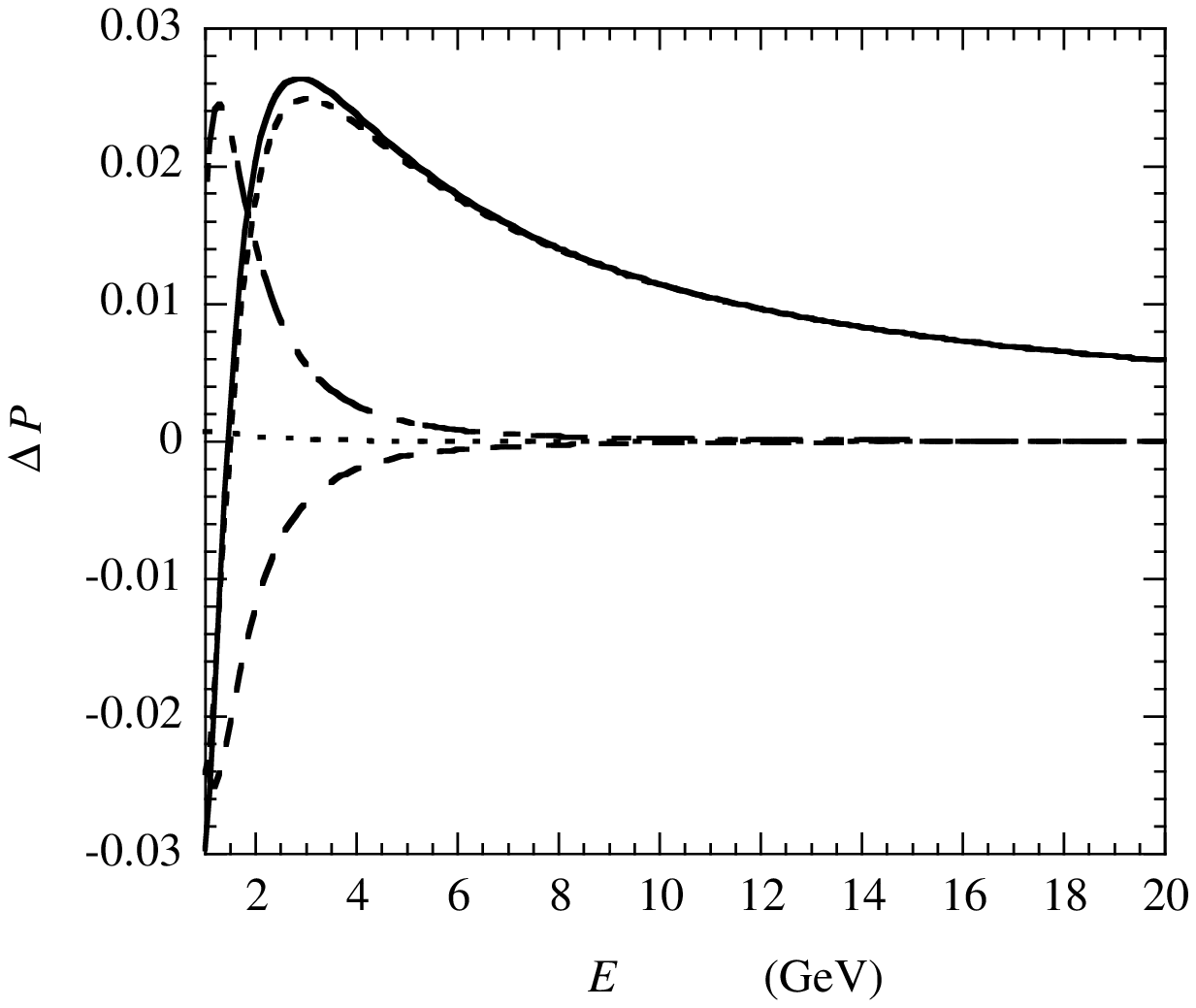}
%\vspace{10mm}
\caption{The CP violating effect in $\nu_{\mu}\to\nu_\tau$ 
oscillation for the neutrino energy range $E=1-20$ GeV at $L=730$ km, with 
the NSNI in $\nu_\mu -\nu_\tau$ sector. The dash-dotted line is the pure 
CP violating effect. The long-dashed, short-dashed and dotted lines are the 
fake CP violating effects due to the ordinary, FCNI and FDNI matter effects, 
respectively. The solid line is the total CP violating effect with the pure 
and fake ones. The parameter values are $|\epsilon|=0.01, \epsilon '=0.02, 
\phi=0, s_{12}=0.54, s_{23}=0.707, s_{13}=0.14, \delta=\pi/2, 
\Delta m^2_{21}=7.3\times 10^{-5}$ ${\rm eV}^2$, and $\Delta m^2_{31}=
2.5\times 10^{-3}{\rm eV}^2$ $(>0)$.}
\end{center}
\end{figure}

\newpage
\begin{figure}[tbp]
\begin{center}
\includegraphics[width=13cm]{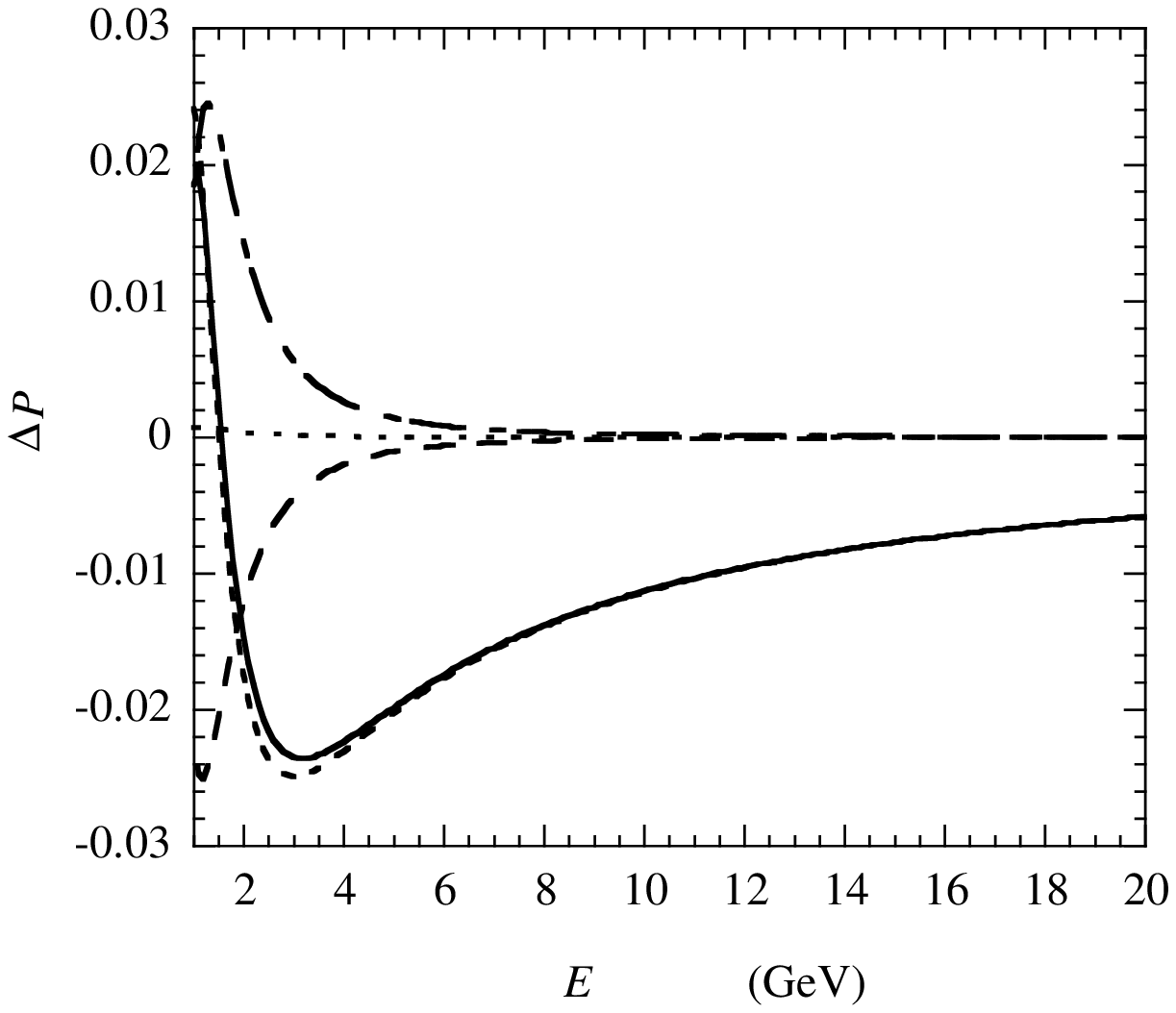}
%\vspace{10mm}
\caption{The CP violating effect in $\nu_\mu\to\nu_\tau$ oscillation 
at $L=730$ km with the same parameter values as in Fig.3 except for 
$\phi=\pi$. The lines represent the same ones as in Fig.3.}
\end{center}
\end{figure}

\newpage
\begin{figure}[tbp]
\begin{center}
\includegraphics[width=13cm]{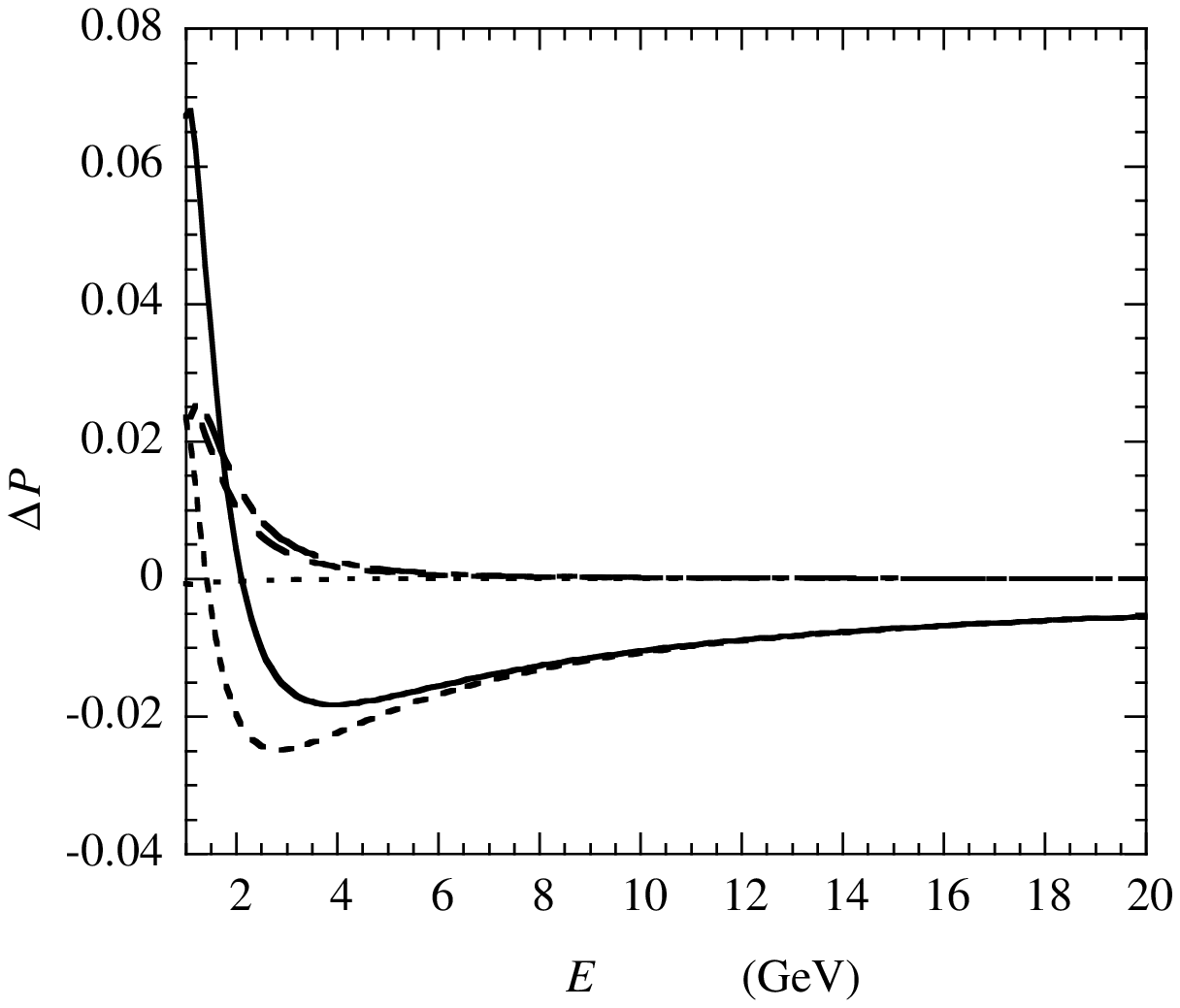}
%\vspace{10mm}
\caption{The CP violating effect in $\nu_\mu\to\nu_\tau$ oscillation 
at $L=730$ km with the same parameter values as in Fig.3 except for 
$\Delta m^2_{31}<0$. The lines represent the same ones as in Fig.3.}
\end{center}
\end{figure}

\newpage
\begin{figure}[tbp]
\begin{center}
\includegraphics[width=13cm]{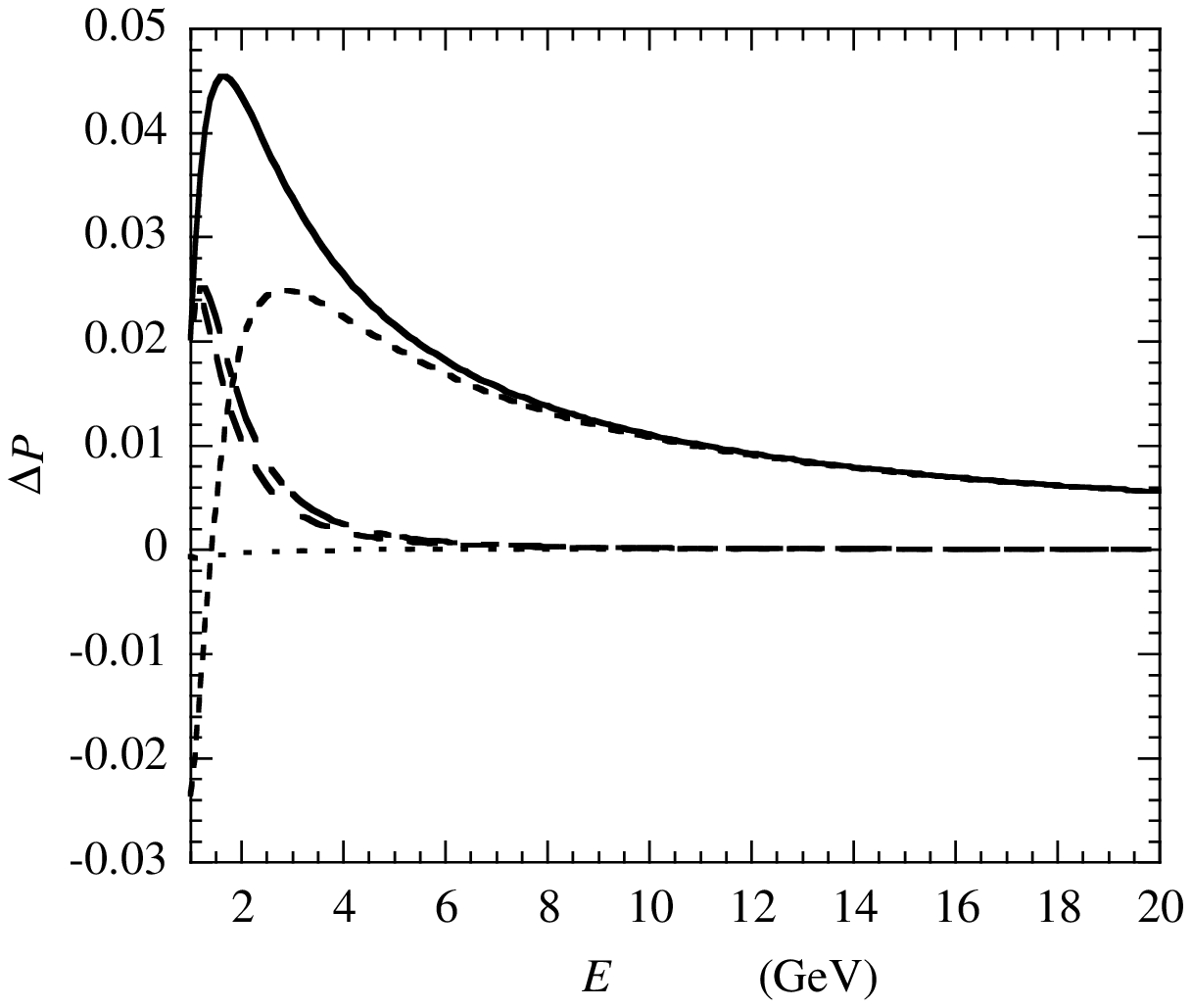}
%\vspace{10mm}
\caption{The CP violating effect in $\nu_\mu\to\nu_\tau$ oscillation 
at $L=730$ km with the same parameter values as in Fig.3 except for 
$\phi=\pi$ and $\Delta m^2_{31}<0$. The lines represent the same ones as 
in Fig.3.}
\end{center}
\end{figure}

\newpage
\begin{figure}[tbp]
\begin{center}
\includegraphics[width=13cm]{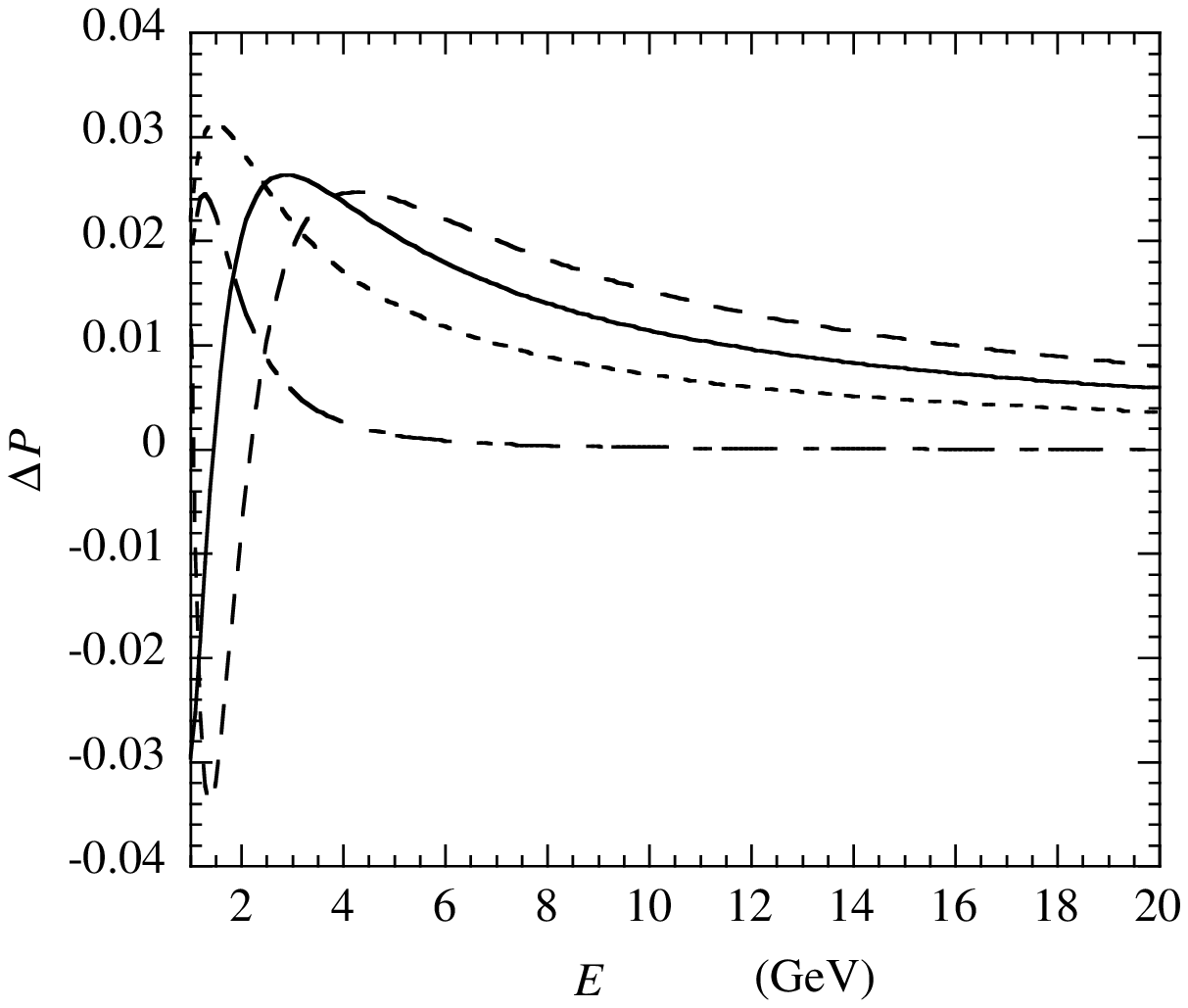}
%\vspace{10mm}
\caption{The dependence on $\Delta m^2_{31}$ of the total CP violating 
effect in $\nu_\mu\to\nu_\tau$ oscillation at $L=730$ km. The short-dashed, 
solid, and long-dashed lines are for $\Delta m^2_{31}=1.5\times 10^{-3}, 
2.5\times 10^{-3}$, and $3.5\times 10^{-3}$ ${\rm eV}^2$, respectively. 
The other mixing parameters are the same as in Fig.3. The dash-dotted line 
represents the pure CP violating effect without the FCNI and FDNI for 
$\Delta m^2_{31}=2.5\times 10^{-3}$ ${\rm eV}^2$.}
\end{center}
\end{figure}

\newpage
\begin{figure}[tbp]
\begin{center}
\includegraphics[width=13cm]{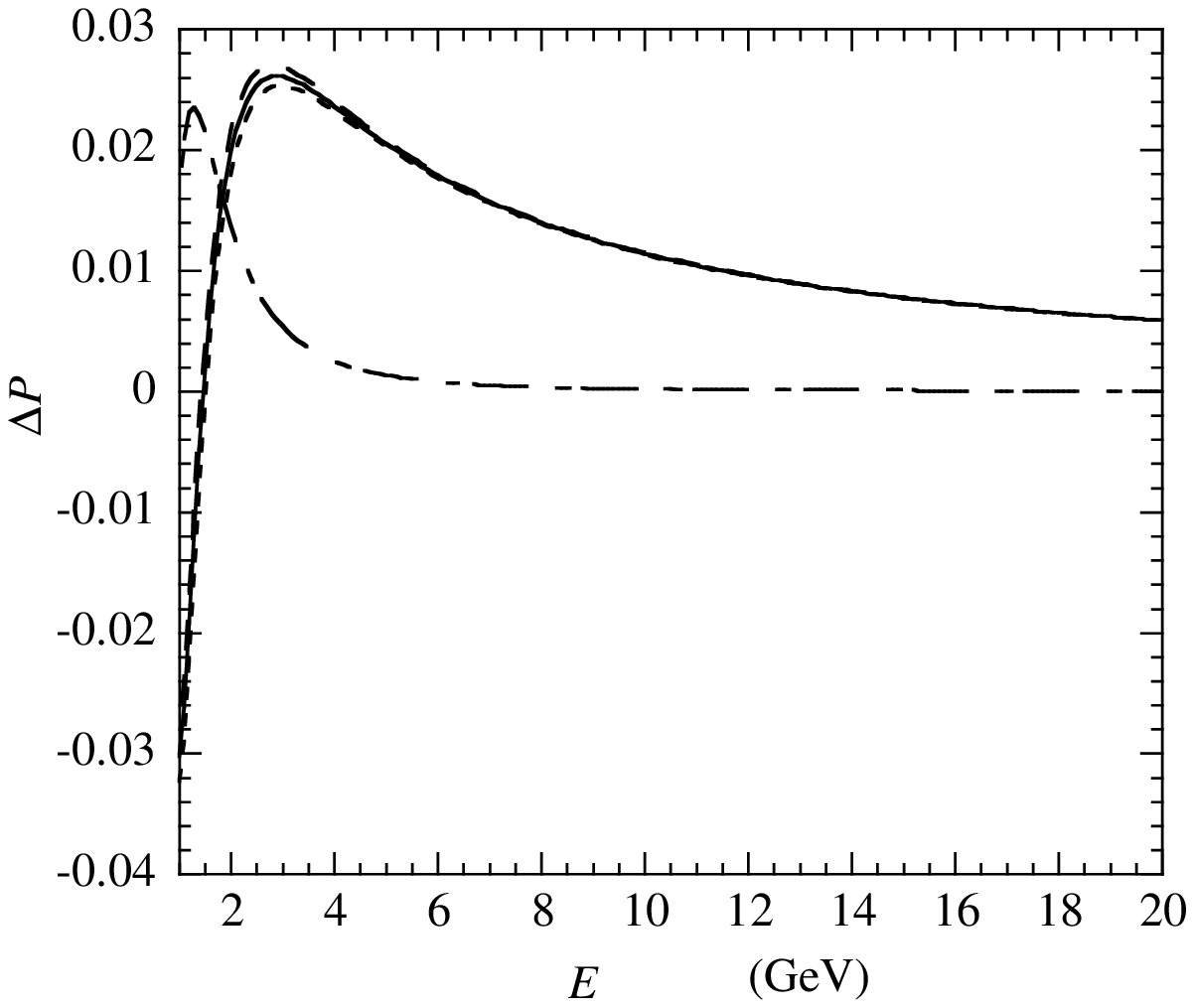}
%\vspace{10mm}
\caption{The dependence on $\Delta m^2_{21}$ of the total CP violating 
effect in $\nu_\mu\to\nu_\tau$ oscillation at $L=730$ km. The short-dashed, 
solid, and long-dashed lines are for $\Delta m^2_{21}=6\times 10^{-5}, 
7\times 10^{-5}$, and $8\times 10^{-5}$ ${\rm eV}^2$, respectively. 
The other mixing parameters are the same as in Fig.3. The dash-dotted line 
represents the pure CP violating effect without the FCNI and FDNI for 
$\Delta m^2_{21}=7\times 10^{-5}$ ${\rm eV}^2$.}
\end{center}
\end{figure}

\newpage
\begin{figure}[tbp]
\begin{center}
\includegraphics[width=13cm]{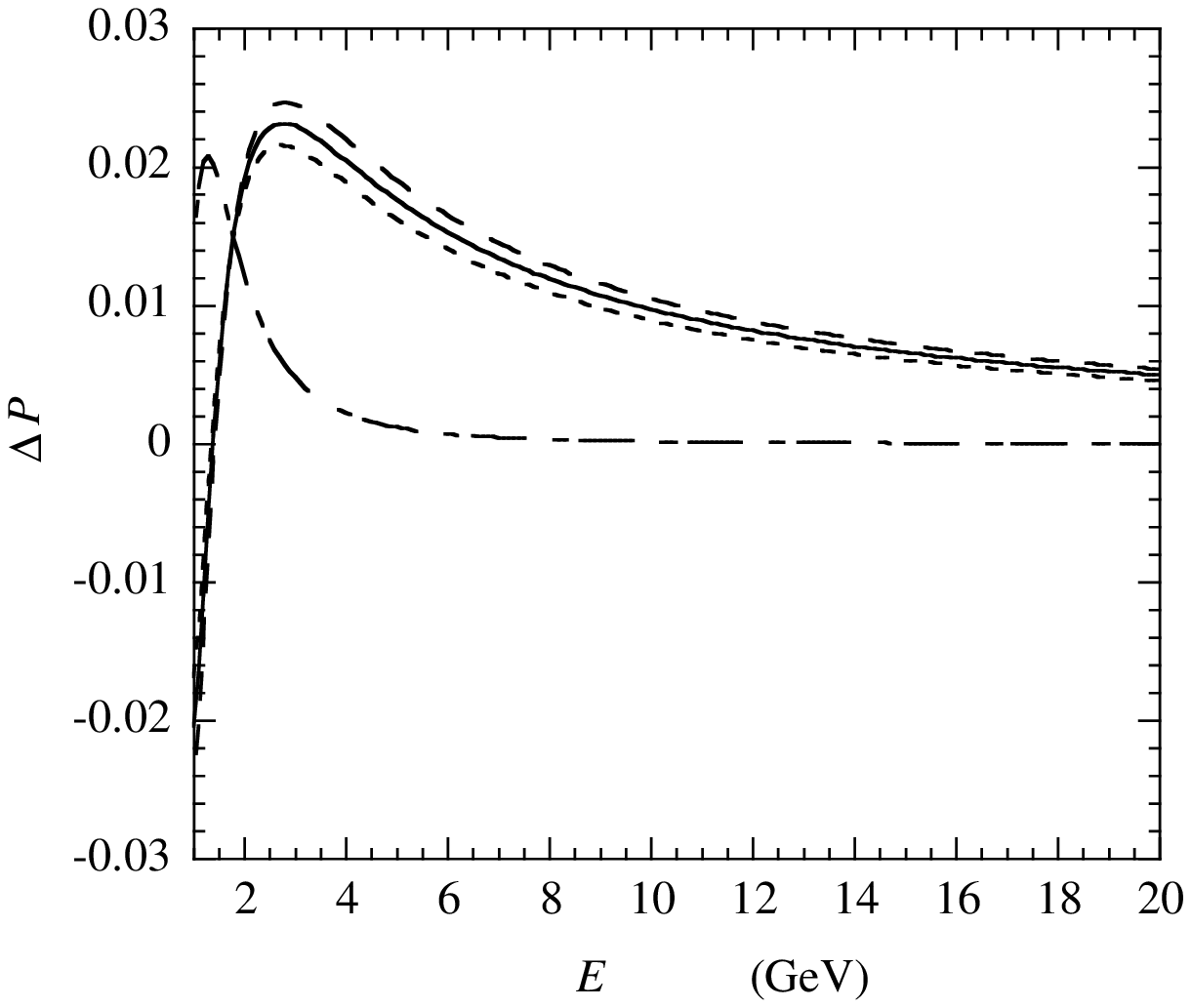}
%\vspace{10mm}
\caption{The dependence on the mixing angle $s_{23}$ of the total CP 
violating effect in $\nu_\mu\to\nu_\tau$ oscillation at $L=730$ km. 
The short-dashed, solid, and long-dashed lines are for $s_{23}=0.55, 0.60$ 
and 0.65, respectively. The other mixing parameters are the same as in Fig.3. 
The dash-dotted line represents the pure CP violating effect without the FCNI 
and FDNI for $s_{23}=0.65$.}
\end{center}
\end{figure}

\newpage
\begin{figure}[tbp]
\begin{center}
\includegraphics[width=13cm]{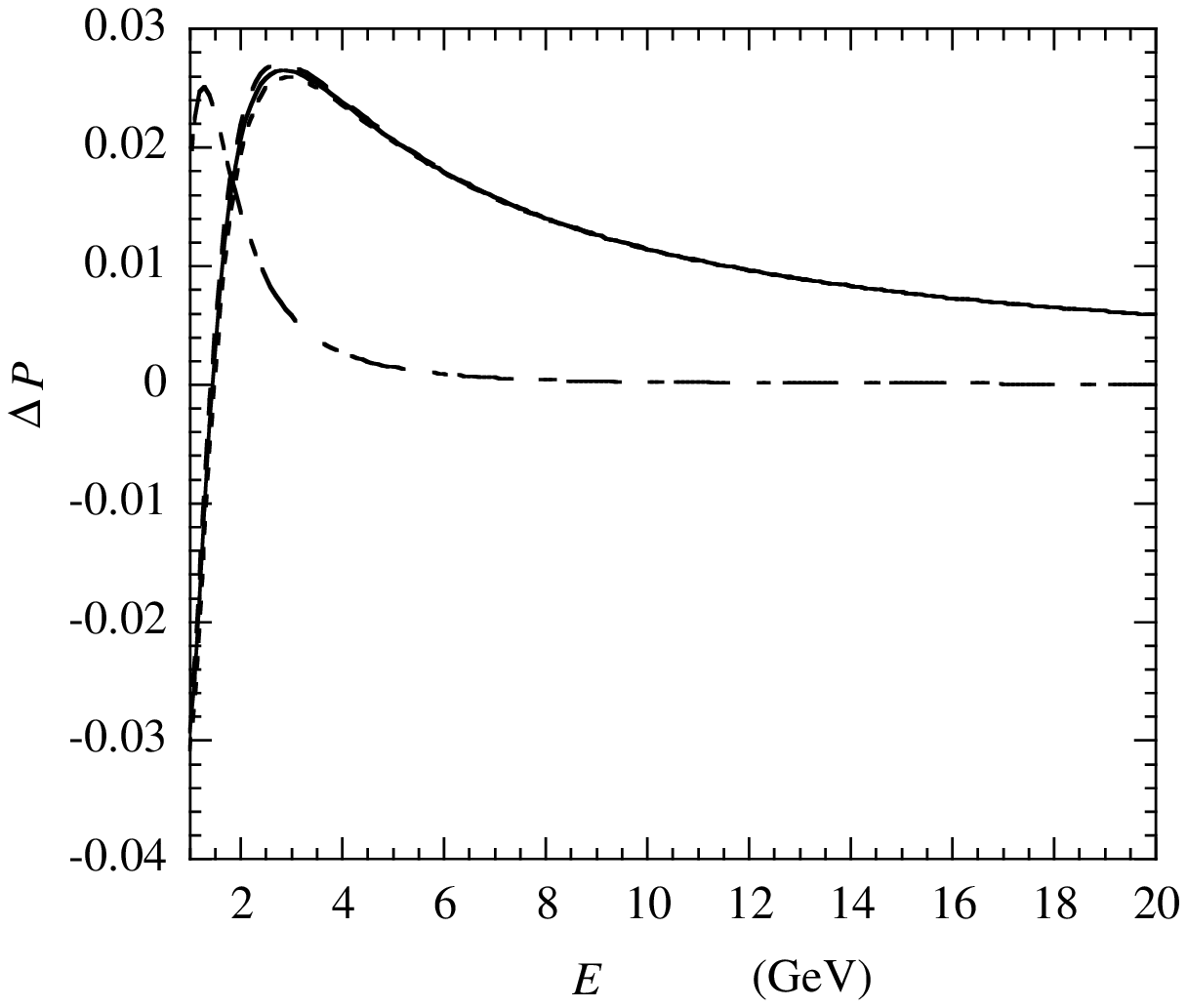}
%\vspace{10mm}
\caption{The dependence on the mixing angle $s_{12}$ of the total CP 
violating effect in $\nu_\mu\to\nu_\tau$ oscillation at $L=730$ km. 
The short-dashed, solid, and long-dashed lines are for $s_{12}=0.50, 0.55$ 
and 0.60, respectively. The other mixing parameters are the same as in Fig.3. 
The dash-dotted line represents the pure CP violating effect without the FCNI 
and FDNI for $s_{12}=0.55$.}
\end{center}
\end{figure}

\newpage
\begin{figure}[tbp]
\begin{center}
\includegraphics[width=13cm]{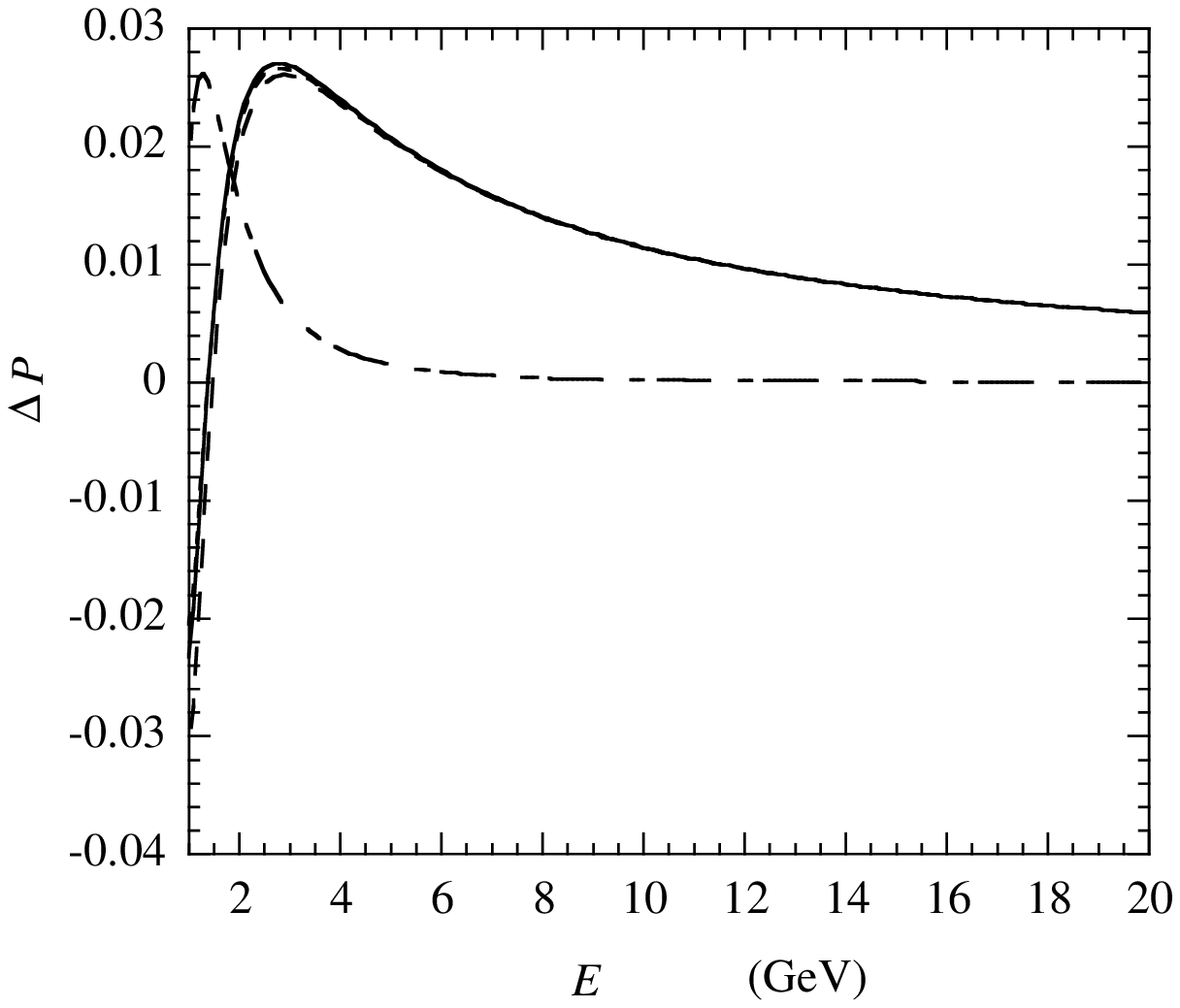}
%\vspace{10mm}
\caption{The dependence on the mixing angle $s_{13}$ of the total CP 
violating effect in $\nu_\mu\to\nu_\tau$ oscillation at $L=730$ km. 
The short-dashed, solid, and long-dashed lines are for $s_{13}=0.05, 0.10$ 
and 0.15, respectively. The other mixing parameters are the same as in Fig.3. 
The dash-dotted line represents the pure CP violating effect without the FCNI 
and FDNI for $s_{13}=0.15$.}
\end{center}
\end{figure}

\newpage
\begin{figure}[tbp]
\begin{center}
\includegraphics[width=13cm]{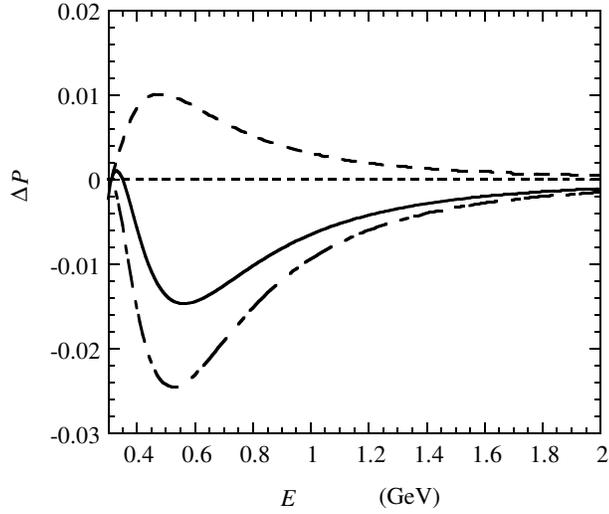}
%\vspace{10mm}
\caption{The CP violating effect in $\nu_\mu\to\nu_e$ 
oscillation for the neutrino energy range $E=0.3-2$ GeV at $L=300$ km, with 
the NSNI in $\nu_e -\nu_\mu$ sector. The dash-dotted line is the pure 
CP violating effect. The long-dashed and short-dashed lines are the 
fake CP violating effects due to the ordinary and FCNI matter effects, 
respectively. The solid line is the total CP violating effect with the pure 
and fake ones. The parameter values are $|\eta|=7\times 10^{-5}, \eta '=0.01, 
\gamma=\pi/2, s_{12}=0.54, s_{23}=0.707, s_{13}=0.14, \delta=\pi/2, 
\Delta m^2_{21}=7.3\times 10^{-5}$ ${\rm eV}^2$, and $\Delta m^2_{31}=
2.5\times 10^{-3}{\rm eV}^2$ $(>0)$.}
\end{center}
\end{figure}

\end{document}